\documentclass[a4paper,10pt]{article}
\usepackage[english]{babel}
\usepackage[T1]{fontenc}
\usepackage[utf8]{inputenc}
\usepackage{amssymb}
\usepackage{physics}
\usepackage{amsmath}
\usepackage{amsthm}

\usepackage{graphicx}
\usepackage[unicode]{hyperref}
\usepackage{tikz}
\usetikzlibrary{arrows.meta,decorations.markings}
\usepackage{color}
\usepackage{genyoungtabtikz}
\usepackage[title]{appendix}
\usepackage{indentfirst}
\usepackage{bm}
\usepackage{xhfill}
\usepackage{bbm}
\usepackage{multirow}
\usepackage{array}
\usepackage[centertableaux]{ytableau}
\usepackage{multicol}
\usepackage[shortlabels]{enumitem}
\usepackage{forest}
\usepackage{float}
\usepackage{mathrsfs}

\usetikzlibrary{decorations.pathreplacing}

\usepackage{ytableau}

\usepackage{xcolor}
\usepackage{tikz}
\usepackage{graphicx}

\usetikzlibrary{
    arrows.meta,
    decorations.markings
}

\definecolor{activeblue}{RGB}{0,85,210}
\definecolor{inactivered}{RGB}{220,35,35}
\definecolor{contourviolet}{RGB}{165,0,200}

\hypersetup{
    colorlinks=true,
    linkcolor=blue,
    filecolor=magenta,      
    urlcolor=cyan,
}
 
\usepackage{amsthm}

\theoremstyle{definition}

\usepackage[nottoc]{tocbibind}

\usepackage{authblk}

\title{\textbf{Complexified Virasoro Zero-Mode Flow and the Logarithmic Graviton at the Chiral Point}}

\author[1,2]{\textbf{Yannick Mvondo-She}}
\affil[1]{National Institute of Theoretical and Computational Sciences, Private Bag X1, Matieland, South Africa}
\affil[1]{\texttt{yannick.mvondo-she@nithecs.ac.za}}
\affil[2]{Department of Physics, University of Pretoria, Private Bag X20, Hatfield 0028, \hspace{1cm} South Africa}

\date{}

\begin{document}

\maketitle

\begin{abstract}
We investigate the relation between Virasoro zero-mode evolution and
the bulk logarithmic graviton in topologically massive gravity at the
chiral point. By introducing additive coordinates adapted to the
global time and radial dependence of the logarithmic mode, we obtain
the expression
\[
y(s,\bar s)
=
-s+\ln2
-\ln\!\left(1+e^{-(s+\bar s)}\right),
\qquad
s=\rho+i\tau,
\quad
\bar s=\rho-i\tau .
\]
This relation provides a direct bulk realization of the additive
coordinates previously associated with the joint Virasoro zero-mode
evolution. Near the AdS$_3$ boundary, the logarithmic coefficient
becomes linear in $s$ up to an additive constant and exponentially
suppressed corrections, while its finite-radius dependence is
governed by the diagonal combination $s+\bar s=2\rho$. Upon
complexification of the radial coordinate, this combination also
organizes the periodic branch structure underlying the previously
established bulk monodromy of the logarithmic graviton. The same
diagonal coordinate controls the nilpotent part of the Virasoro
zero-mode representation, providing a common additive-coordinate
description of the representation-theoretic Jordan structure and the
radial analytic structure of the bulk solution. The corresponding
operations remain distinct: finite zero-mode evolution is algebraic,
whereas bulk monodromy arises from analytic continuation and winding
of the multivalued gravitational solution.
\end{abstract}

\tableofcontents

\section{Introduction}
\label{sec:introduction}

Three-dimensional topologically massive gravity (TMG) provides a
particularly economical framework in which local gravitational
degrees of freedom, asymptotic conformal symmetry, and holographic
structure coexist. The theory is obtained by augmenting the
Einstein--Hilbert action with negative cosmological constant by a
gravitational Chern--Simons term
\cite{Deser:1981wh,Deser:1982vy}. On an AdS$_3$ background of radius
$\ell$, the resulting theory contains a propagating massive graviton,
with its linearized spectrum governed by the dimensionless parameter
$\mu\ell$, where $\mu$ denotes the topological mass.

The critical value
\begin{equation}
\mu\ell=1
\end{equation}
acquired particular significance following the proposal of
Ref.~\cite{Li:2008dq}. At this \emph{chiral point}, the massive and
left-moving branches of the linearized theory become degenerate, and
it was argued that, with Brown--Henneaux boundary conditions, the
problematic bulk excitations could be removed, leaving a purely
right-moving theory known as chiral gravity. The proposal stimulated
a detailed investigation of the spectrum and boundary conditions of
critical TMG
\cite{Carlip:2008jk,Giribet:2008bw,Grumiller:2008qz}. In particular,
the degeneracy of the linearized operator was shown to admit an
additional logarithmic solution with nontrivial bulk dynamics
\cite{Grumiller:2008qz}. Subsequent analyses developed the resulting
logarithmic sector and its holographic interpretation
\cite{Grumiller:2008pr,Grumiller:2008es,Grumiller:2009mw,
Ertl:2010dh,Gaberdiel:2010xv,Grumiller:2010rm,
Grumiller:2010tj,Grumiller:2013at}.

The physical theory obtained at the critical point depends crucially
on the asymptotic boundary conditions. With strict Brown--Henneaux
asymptotically AdS$_3$ boundary conditions
\cite{Brown:1986nw}, the logarithmic modes are excluded and one
obtains the chiral-gravity sector, whose quantum and path-integral
properties have motivated further developments in three-dimensional
gravity \cite{Maloney:2015ina,Eberhardt:2022wlc}. The present work
instead concerns the theory obtained when the boundary conditions are
relaxed sufficiently to retain the logarithmic solution
\cite{Grumiller:2008qz}. This theory, commonly referred to as
\emph{log gravity} \cite{Maloney:2009ck}, admits logarithmic
asymptotic behavior and is naturally associated holographically with
a logarithmic conformal field theory. Holographic renormalization at
the critical point reproduces the characteristic logarithmic
correlators and the new anomaly of the dual LCFT
\cite{Skenderis:2009nt}.

Logarithmic conformal field theories are distinguished by
non-diagonalizable actions of the conformal generators and the
resulting indecomposable representation structure
\cite{Gurarie:1993xq,Flohr:2001zs}. Their connection with AdS/CFT
\cite{Kogan:1999bn} and, more specifically, with
AdS$_3$/LCFT$_2$ holography \cite{Myung:1999nd} provides the broader
representation-theoretic setting for the logarithmic sector of
critical TMG. Rank-two Jordan cells, their extension structure, and
the propagation of logarithmic mixing through descendant modules are
standard features of this framework
\cite{Kytola:2009ax}. In critical TMG, these structures are realized
in the bulk by the degeneracy of the massive and left-moving graviton
branches and the associated logarithmic partner.

In the Grumiller--Johansson normalization, the logarithmic graviton
takes the form
\begin{equation}
\psi^{\rm new}_{\mu\nu}
=
y(\tau,\rho)\psi^L_{\mu\nu},
\qquad
y(\tau,\rho)
=
-i\tau-\ln\cosh\rho.
\label{eq:intro_GJ_mode}
\end{equation}
Its logarithmic structure admits complementary algebraic and
complex-analytic descriptions. The Virasoro zero modes act
non-diagonalizably on the logarithmic pair, while complexification of
the radial coordinate exposes the multivalued analytic structure of
the bulk coefficient $y$. Bulk radial monodromy and its extension to
the global descendant sector were developed in
Refs.~\cite{Mvondo-She:2026egr,Mvondo-She:2026eez}, whereas the joint
action of the commuting Virasoro zero modes and its additive
parametrization were studied in Ref.~\cite{Mvondo-She:2026vmi}.

For the logarithmic pair in the bulk-inherited normalization, the
finite joint zero-mode evolution may be written as
\begin{equation}
U(s,\bar s)
=
e^{-hs-\bar h\bar s}
\left[
\mathbf 1-\frac12(s+\bar s)N
\right],
\qquad
N^2=0,
\label{eq:intro_joint_flow}
\end{equation}
where $(s,\bar s)$ are additive parameters for the commuting
zero-mode actions. In Ref.~\cite{Mvondo-She:2026vmi}, these variables
were also related to the multiplicative grading coordinates
$q=e^{-s}$ and $\bar q=e^{-\bar s}$. The question addressed here is
whether the same additive coordinates admit a direct realization in
the spacetime dependence of the bulk logarithmic graviton.

To this end, we introduce the bulk parametrization
\begin{equation}
s=\rho+i\tau,
\qquad
\bar s=\rho-i\tau,
\label{eq:intro_bulk_parametrization}
\end{equation}
so that
\begin{equation}
s+\bar s=2\rho,
\qquad
s-\bar s=2i\tau.
\label{eq:intro_sum_difference}
\end{equation}
Substitution into Eq.~\eqref{eq:intro_GJ_mode} gives the identity
\begin{equation}
y(s,\bar s)
=
-s+\ln2
-\ln\!\left(1+e^{-(s+\bar s)}\right).
\label{eq:intro_central_identity}
\end{equation}
This relation is the central result of the present construction. It
provides a direct bulk realization of the additive zero-mode
coordinates and separates the logarithmic coefficient into a term
linear in $s$ and a radial contribution governed by the
diagonal coordinate $s+\bar s$.

Two consequences follow immediately. First, near the AdS$_3$
boundary,
\begin{equation}
y(s,\bar s)
=
-s+\ln2+O(e^{-2\rho}),
\qquad
\rho\rightarrow\infty,
\label{eq:intro_asymptotic_s}
\end{equation}
so that $s$ governs the leading complex-linear dependence of the
logarithmic graviton. Second, upon complexification of the radial
coordinate, the logarithmic term in
Eq.~\eqref{eq:intro_central_identity} has the periodic branch locus
\begin{equation}
s+\bar s=(2m+1)\pi i,
\qquad
m\in\mathbb Z.
\label{eq:intro_branch_locus}
\end{equation}
Since $s+\bar s=2\rho$, this gives an additive-coordinate
representation of the radial analytic structure underlying the
previously established bulk monodromy. Under the exponential map
\begin{equation}
r=e^\rho=e^{(s+\bar s)/2},
\end{equation}
the periodic branch sequence projects to $r=\pm i$. The remaining
$r$-plane branch point $r=0$ has no finite preimage in the additive
coordinate and instead corresponds to the asymptotic end
$\operatorname{Re}(s+\bar s)\rightarrow-\infty$.

Taken together, these results show that the additive coordinates
provide a common geometric framework for several structures that
appear separately in the conventional formulation of critical TMG:
the spacetime dependence of the logarithmic graviton, the nilpotent
part of the Virasoro zero-mode representation, the radial analytic
structure, and the associated unipotent bulk monodromy. In this
description, the diagonal and anti-diagonal directions separate the
nilpotent and semisimple parts of the zero-mode parameter
space, while the diagonal coordinate $s+\bar s$ also organizes the
radial analytic structure of the bulk logarithmic coefficient. The
same coordinate framework thus relates the zero-mode Jordan structure
to the covering-space geometry of the radial continuation, while
retaining the different mathematical origins of the corresponding
transformations: finite zero-mode evolution is determined by additive
endpoint data, whereas bulk monodromy arises from analytic
continuation and winding around the branch structure of the
multivalued gravitational solution.

The remainder of the paper is organized as follows.
Section~\ref{sec:log_graviton_zero_modes} recalls the minimal
logarithmic-graviton and Virasoro zero-mode structure required for the
analysis. Section~\ref{sec:additive_coordinates_bulk} introduces the
bulk parametrization of $(s,\bar s)$ and derives the relation
\eqref{eq:intro_central_identity} and its asymptotic limit.
Section~\ref{sec:radial_complexification_monodromy} determines the
radial branch structure in additive-coordinate space and relates it
to the exponential radial description.
Section~\ref{sec:zero_mode_vs_bulk_monodromy} compares the diagonal
and anti-diagonal directions, the corresponding continuations, and
the relation to the previously established monodromy constructions.
Finally, Sec.~\ref{sec:discussion} summarizes the resulting picture
and discusses possible extensions of the additive-coordinate
construction.

\section{Logarithmic Graviton and Virasoro Zero-Mode Structure}
\label{sec:log_graviton_zero_modes}

We begin by recalling the minimal bulk and representation-theoretic
structure required for the analysis. At the chiral point of
topologically massive gravity, the massive and left-moving graviton
branches become degenerate and an additional logarithmic solution
appears. The resulting logarithmic pair carries a rank-two
indecomposable action of the commuting conformal zero modes
$L_0$ and $\bar L_0$. The corresponding zero-mode structure and its
additive parametrization were developed in
Ref.~\cite{Mvondo-She:2026vmi}; here we retain only the ingredients
needed to relate those additive variables directly to the bulk
logarithmic graviton.

\subsection{Critical TMG and the logarithmic graviton}
\label{subsec:critical_tmg_log_graviton}

Topologically massive gravity (TMG) with a negative cosmological
constant is governed by
\begin{equation}
G_{\mu\nu}
-\frac{1}{\ell^2}g_{\mu\nu}
+\frac{1}{\mu}C_{\mu\nu}
=0,
\label{eq:tmg_eom}
\end{equation}
where $\ell$ is the AdS$_3$ radius, $\mu$ is the topological mass
parameter, and $C_{\mu\nu}$ is the Cotton tensor. For linearized
perturbations about AdS$_3$, the field equation admits the
factorization
\begin{equation}
\left(
\mathcal D^{L}
\mathcal D^{R}
\mathcal D^{M} h
\right)_{\mu\nu}
=0.
\label{eq:tmg_factorization}
\end{equation}
At the chiral point
\begin{equation}
\mu\ell=1,
\label{eq:chiral_point}
\end{equation}
the massive and left-moving operators degenerate,
$\mathcal D^{M}=\mathcal D^{L}$, so that
\begin{equation}
\left(
\mathcal D^{R}
(\mathcal D^{L})^2 h
\right)_{\mu\nu}
=0.
\label{eq:critical_factorization}
\end{equation}
In addition to the ordinary left-moving mode
\begin{equation}
\mathcal D^{L}\psi^L_{\mu\nu}=0,
\end{equation}
the repeated factor admits a generalized logarithmic solution
$\psi^{\rm new}_{\mu\nu}$ satisfying
\begin{equation}
(\mathcal D^{L})^2\psi^{\rm new}_{\mu\nu}=0,
\qquad
\mathcal D^{L}\psi^{\rm new}_{\mu\nu}\neq0.
\label{eq:log_generalized_kernel}
\end{equation}

In global AdS$_3$ coordinates $(\tau,\rho,\phi)$, the logarithmic
graviton may be written in the Grumiller--Johansson normalization as
\begin{equation}
\psi^{\rm new}_{\mu\nu}
=
y(\tau,\rho)\psi^L_{\mu\nu},
\qquad
y(\tau,\rho)
=
-i\tau-\ln\cosh\rho,
\label{eq:gj_log_mode}
\end{equation}
with left-moving conformal weights
\begin{equation}
(h,\bar h)=(2,0).
\label{eq:critical_weights}
\end{equation}
This normalization will be retained throughout the paper. The
logarithmic coefficient $y(\tau,\rho)$ is the only bulk input needed
for the construction below. Its relation to the additive zero-mode
coordinates will be derived directly in
Sec.~\ref{sec:additive_coordinates_bulk}.

\subsection{Joint action of the conformal zero modes}
\label{subsec:joint_zero_modes}

The Virasoro zero-mode structure of the logarithmic pair was analyzed
in detail in Ref.~\cite{Mvondo-She:2026vmi}. We recall here only the
normalization and finite joint evolution required below.

In global light-cone coordinates
\begin{equation}
u=\tau+\phi,
\qquad
v=\tau-\phi,
\end{equation}
the conformal zero modes are
\begin{equation}
L_0=i\partial_u,
\qquad
\bar L_0=i\partial_v.
\label{eq:global_zero_modes}
\end{equation}
For the Grumiller--Johansson logarithmic coefficient
\begin{equation}
y(\tau,\rho)=-i\tau-\ln\cosh\rho,
\end{equation}
one has
\begin{equation}
L_0y=\bar L_0y=\frac12.
\label{eq:zero_modes_on_y}
\end{equation}
Consequently, if
\begin{equation}
L_0\psi^L_{\mu\nu}=h\psi^L_{\mu\nu},
\qquad
\bar L_0\psi^L_{\mu\nu}=\bar h\psi^L_{\mu\nu},
\end{equation}
then
\begin{subequations}
\begin{align}
L_0\psi^{\rm new}_{\mu\nu}
&=
h\psi^{\rm new}_{\mu\nu}
+\frac12\psi^L_{\mu\nu},
\\
\bar L_0\psi^{\rm new}_{\mu\nu}
&=
\bar h\psi^{\rm new}_{\mu\nu}
+\frac12\psi^L_{\mu\nu}.
\end{align}
\label{eq:zero_mode_log_pair}
\end{subequations}
The factor $1/2$ is fixed by the bulk normalization and will be
retained throughout.

Introducing the nilpotent operator
\begin{equation}
N\psi^L_{\mu\nu}=0,
\qquad
N\psi^{\rm new}_{\mu\nu}=\psi^L_{\mu\nu},
\qquad
N^2=0,
\label{eq:nilpotent_N}
\end{equation}
the zero modes take the compact form
\begin{equation}
L_0=h\mathbf 1+\frac12N,
\qquad
\bar L_0=\bar h\mathbf 1+\frac12N.
\label{eq:individual_zero_mode_jordan}
\end{equation}
In particular,
\begin{equation}
L_0+\bar L_0
=
(h+\bar h)\mathbf 1+N,
\qquad
L_0-\bar L_0
=
(h-\bar h)\mathbf 1,
\label{eq:sum_difference_zero_modes}
\end{equation}
so that the nilpotent contribution is carried by the sum of the two
zero modes.

For additive parameters $(s,\bar s)$, the corresponding finite joint
evolution is
\begin{equation}
U(s,\bar s)
:=
e^{-sL_0-\bar s\bar L_0}.
\label{eq:joint_zero_mode_evolution}
\end{equation}
Since $N^2=0$, Eq.~\eqref{eq:individual_zero_mode_jordan} gives
\begin{equation}
U(s,\bar s)
=
e^{-hs-\bar h\bar s}
\left[
\mathbf 1-\frac12(s+\bar s)N
\right].
\label{eq:finite_joint_zero_mode_flow}
\end{equation}
Thus the nilpotent part of the joint evolution depends only on
\begin{equation}
\sigma:=s+\bar s.
\label{eq:sigma_definition}
\end{equation}
This is the only feature of the finite zero-mode evolution that will
be needed for the bulk construction below.

\subsection{Additive Virasoro zero-mode coordinates}
\label{subsec:additive_zero_mode_coordinates}

The parameters $(s,\bar s)$ in
Eq.~\eqref{eq:joint_zero_mode_evolution} are additive coordinates for
the commuting zero-mode evolution. Since
\begin{equation}
[L_0,\bar L_0]=0,
\end{equation}
they satisfy the composition law
\begin{equation}
U(s_1,\bar s_1)\,
U(s_2,\bar s_2)
=
U(s_1+s_2,\bar s_1+\bar s_2).
\label{eq:additive_group_law}
\end{equation}
Their exponentials define the corresponding multiplicative conformal
grading variables,
\begin{equation}
q=e^{-s},
\qquad
\bar q=e^{-\bar s},
\label{eq:q_from_additive_coordinates}
\end{equation}
or, on a chosen logarithmic branch,
\begin{equation}
s=-\log q,
\qquad
\bar s=-\log\bar q.
\label{eq:additive_from_q}
\end{equation}
The relation between these additive coordinates and logarithmic
correlators and partition-function grading was developed in
Ref.~\cite{Mvondo-She:2026vmi}.

On a real section, it is convenient to write
\begin{equation}
s=u+i\theta,
\qquad
\bar s=u-i\theta,
\qquad
u,\theta\in\mathbb R,
\label{eq:u_theta_parametrization}
\end{equation}
so that
\begin{equation}
s+\bar s=2u,
\qquad
s-\bar s=2i\theta.
\label{eq:sum_difference_u_theta}
\end{equation}
Together with Eq.~\eqref{eq:finite_joint_zero_mode_flow}, this shows
that the nilpotent part of the joint evolution is controlled by the
common additive direction $u=(s+\bar s)/2$.

The question addressed in the remainder of this work is whether the
same additive coordinates admit a direct realization in the bulk
logarithmic graviton. We now turn to this question.

\section{Additive Zero-Mode Coordinates in the Bulk Logarithmic Graviton}
\label{sec:additive_coordinates_bulk}

We now construct the bulk realization of these additive coordinates
and determine how the Grumiller--Johansson coefficient is
organized in the resulting $(s,\bar s)$ coordinate space.

\subsection{Bulk parametrization of $(s,\bar s)$}
\label{subsec:bulk_parametrization}

Motivated by the real-section parametrization
$s=u+i\theta$, $\bar s=u-i\theta$ \cite{Mvondo-She:2026vmi}, we introduce
\begin{equation}
s=\rho+i\tau,
\qquad
\bar s=\rho-i\tau,
\label{eq:bulk_s_sbar}
\end{equation}
where $\rho$ and $\tau$ are the radial and global-time coordinates of
AdS$_3$. The bulk reality condition used here should be distinguished from the real-section parametrization of the zero-mode parameter space in
Ref.~\cite{Mvondo-She:2026vmi}. There, $s=u+i\theta$ and $\bar s=u-i\theta$, with $u,\theta\in\mathbb R$, provide additive coordinates adapted to the conformal zero-mode evolution and its grading variables. In the
present construction the same additive-coordinate form is instead given a direct bulk interpretation, with $\rho$ the radial coordinate and $\tau$ the real global time of Lorentzian AdS$_3$. The formal parallel between $(u,\theta)$ and $(\rho,\tau)$ therefore concerns two realizations of the same additive parameter space rather than an identification of their physical roles. The radial variable $\rho$ will subsequently be analytically continued away from the physical Lorentzian section in order to study the branch structure of the logarithmic graviton.

On the physical Lorentzian section,
$\rho,\tau\in\mathbb R$, one has $\bar s=s^{*}$, and
\begin{equation}
\rho=\frac{s+\bar s}{2},
\qquad
\tau=\frac{s-\bar s}{2i}.
\label{eq:rho_tau_from_s}
\end{equation}
Equivalently,
\begin{equation}
s+\bar s=2\rho,
\qquad
s-\bar s=2i\tau.
\label{eq:sum_difference_bulk}
\end{equation}
Thus radial and temporal variations correspond, respectively, to the
diagonal and anti-diagonal directions of the $(s,\bar s)$ coordinate
space.

We now apply these coordinates to the logarithmic coefficient
\begin{equation}
y(\tau,\rho)
=
-i\tau-\ln\cosh\rho.
\label{eq:y_before_s_rewriting}
\end{equation}
As shown in the next subsection, its rewriting in terms of
$(s,\bar s)$ provides the direct bulk realization of the additive
coordinates sought here.

\subsection{Logarithmic coefficient in $(s,\bar s)$}
\label{subsec:exact_log_coefficient}

The bulk parametrization in Eq.~\eqref{eq:bulk_s_sbar} allows the
Grumiller--Johansson coefficient to be expressed directly in terms of
the additive coordinates. Using Eq. (\ref{eq:rho_tau_from_s}) in Eq. (\ref{eq:y_before_s_rewriting}) gives
\begin{equation}
y(s,\bar s)
=
-\frac{s-\bar s}{2}
-
\ln\cosh\!\left(\frac{s+\bar s}{2}\right).
\label{eq:y_s_sbar_first_form}
\end{equation}
The identity
\begin{equation}
\cosh\rho
=
\frac{e^\rho}{2}
\left(1+e^{-2\rho}\right)
\label{eq:cosh_factorization}
\end{equation}
implies, on the physical branch,
\begin{equation}
\ln\cosh\rho
=
\rho-\ln2
+
\ln\!\left(1+e^{-2\rho}\right).
\label{eq:log_cosh_exact}
\end{equation}
Since $\rho=(s+\bar s)/2$, substitution into
Eq.~\eqref{eq:y_s_sbar_first_form} yields
\begin{subequations}    
\begin{align}
y(s,\bar s)
&=
-\frac{s-\bar s}{2}
-\frac{s+\bar s}{2}
+\ln2
-\ln\!\left(1+e^{-(s+\bar s)}\right)
\\
&=
-s+\ln2
-\ln\!\left(1+e^{-(s+\bar s)}\right)
.
\label{eq:y_exact_s_sbar}
\end{align}
\end{subequations}
Equation~\eqref{eq:y_exact_s_sbar} is the central relation between
the additive Virasoro zero-mode coordinates and the spacetime
dependence of the bulk logarithmic graviton. Introducing
\begin{equation}
\sigma:=s+\bar s=2\rho,
\end{equation}
the radial correction depends only on $\sigma$, the
same combination that controls the nilpotent contribution to the
joint zero-mode evolution.

\subsection{Asymptotic realization of the $s$-coordinate}
\label{subsec:asymptotic_s_coordinate}

The relation in Eq.~\eqref{eq:y_exact_s_sbar} immediately
determines the asymptotic behavior of the logarithmic coefficient.
On the physical section,
\begin{equation}
s+\bar s=2\rho,
\end{equation}
and therefore, for $\rho\rightarrow+\infty$,
\begin{equation}
\ln\!\left(1+e^{-(s+\bar s)}\right)
=
\ln\!\left(1+e^{-2\rho}\right)
=
e^{-2\rho}+O(e^{-4\rho}).
\label{eq:finite_radius_asymptotic}
\end{equation}
Since $s=\rho+i\tau$, the temporal and radial dependence combine
as
\begin{equation}
y(s,\bar s)
=
-i\tau-\ln\cosh\rho
=
-(\rho+i\tau)+\ln2+O(e^{-2\rho})
=
-s+\ln2+O(e^{-2\rho}),
\qquad
\rho\rightarrow+\infty.
\label{eq:bulk_linear_combination}
\end{equation}
Thus, under the bulk parametrization introduced above, the additive
coordinate $s$, previously introduced as a parameter of Virasoro
zero-mode evolution and related to the multiplicative grading variable
$q=e^{-s}$, also governs the leading complex-linear dependence of the
bulk logarithmic graviton near the AdS$_3$ boundary.

The constant $\ln2$ reflects the standard freedom to shift a logarithmic partner by a multiple of its ordinary partner and does not affect the rank-two nilpotent structure. Working in the original Grumiller--Johansson normalization, we next examine the analytic structure of this radial contribution under complexification of the radial coordinate.

\section{Radial Analytic Structure in Additive Coordinates}
\label{sec:radial_complexification_monodromy}

The representation Eq.~\eqref{eq:y_exact_s_sbar} contains a  logarithmic term depending only on $s+\bar s=2\rho$. Bulk radial monodromy of the logarithmic graviton has been studied previously in
Refs.~\cite{Mvondo-She:2026egr,Mvondo-She:2026eez}. Here we determine how
the associated branch structure is represented in the additive coordinates $(s,\bar s)$.

\subsection{Complexified radial direction}
\label{subsec:radial_complexification}

We complexify the radial coordinate while keeping the global time
real,
\begin{equation}
\rho\in\mathbb C,
\qquad
\tau\in\mathbb R.
\label{eq:rho_complex_tau_real}
\end{equation}
The additive coordinates in Eq. (\ref{eq:bulk_s_sbar}) remain, therefore satisfying Eq. (\ref{eq:sum_difference_bulk}).

Away from the physical Lorentzian section, $\bar s$ need not equal
$s^{*}$; the bar is retained as the label inherited from the
two-parameter zero-mode description.

At fixed $\tau$, a radial displacement
$\rho\rightarrow\rho+\Delta\rho$ induces
\begin{equation}
(s,\bar s)
\longrightarrow
(s+\Delta\rho,\bar s+\Delta\rho),
\label{eq:diagonal_complex_displacement}
\end{equation}
so radial analytic continuation follows the diagonal direction of the
complexified additive-coordinate space. Equivalently, $\sigma=s+\bar s=2\rho$, varies while $s-\bar s=2i\tau$ remains fixed.

The logarithmic term in Eq.~(\ref{eq:y_exact_s_sbar}) is therefore
multivalued as a function of the complexified radial coordinate, with
its branch structure governed entirely by $\sigma=s+\bar s$. We now
determine the corresponding branch locus.

\subsection{Branch locus in additive-coordinate space}
\label{subsec:branch_locus_additive}

The branch locus is determined by the zeros of the logarithmic
argument,
\begin{equation}
1+e^{-\sigma}=0.
\label{eq:branch_zero_condition_sigma}
\end{equation}
Since $e^{-\sigma}=-1$, the corresponding points in the
$\sigma$-plane are
\begin{equation}
\sigma_m=(2m+1)\pi i,
\qquad
m\in\mathbb Z.
\label{eq:sigma_branch_points}
\end{equation}
Using $\sigma=2\rho$, this gives the periodic sequence
\begin{equation}
\rho_m
=
i\pi\left(m+\frac12\right),
\qquad
m\in\mathbb Z.
\label{eq:rho_branch_points}
\end{equation}
Thus the radial branch structure is represented in additive-coordinate
space by an equally spaced sequence along the imaginary $\sigma$-axis,
with period $2\pi i$. These points coincide with the zeros of
$\cosh\rho$ in the original Grumiller--Johansson coefficient.

The logarithmic character of these points is immediate locally.
Writing
\begin{equation}
\sigma=\sigma_m+\delta\sigma,
\qquad
|\delta\sigma|\ll1,
\end{equation}
and using $e^{-\sigma_m}=-1$, one finds
\begin{equation}
1+e^{-\sigma}
=
1-e^{-\delta\sigma}
=
\delta\sigma+O(\delta\sigma^2).
\end{equation}
Hence
\begin{equation}
-\ln\!\left(1+e^{-\sigma}\right)
=
-\ln(\delta\sigma)+O(\delta\sigma),
\label{eq:local_sigma_logarithm}
\end{equation}
showing that every $\sigma_m$ is a simple logarithmic branch point.

The relation to the exponential radial coordinate used in the
previous monodromy analysis follows from
\begin{equation}
r=e^\rho=e^{(s+\bar s)/2}=e^{\sigma/2}.
\label{eq:r_uniformization}
\end{equation}
Under this map, the periodic additive branch locus projects according
to
\begin{equation}
r_m
=
e^{\sigma_m/2}
=
e^{i\pi(m+1/2)}
=
\begin{cases}
+i, & m\ \text{even},\\[1mm]
-i, & m\ \text{odd}.
\end{cases}
\label{eq:branch_points_map}
\end{equation}
Thus the infinite sequence of finite branch points in the additive
coordinate projects to the two points $r=\pm i$.

The $r$-plane analysis of Ref.~\cite{Mvondo-She:2026eez} also contains the
logarithmic branch point $r=0$. Its representation in the additive
coordinate is different from that of $r=\pm i$. Since the exponential
map $r=e^{\sigma/2}$ does not vanish at finite $\sigma$, the origin has
no finite preimage in the $\sigma$-plane. Rather,
\begin{equation}
r\longrightarrow 0
\qquad\Longleftrightarrow\qquad
\operatorname{Re}\sigma\longrightarrow-\infty.
\label{eq:r_zero_sigma_infinity}
\end{equation}
The additive-coordinate description therefore distinguishes the
three finite branch points of the $r$-plane formulation: the points
$r=\pm i$ lift to the periodic sequence
$\sigma_m=(2m+1)\pi i$, whereas $r=0$ corresponds to an asymptotic
end of the additive-coordinate covering space. In this way, the
exponential map relates the periodic branch structure obtained here
to the previously established winding-number description of bulk
radial monodromy.

\section{Zero-Mode Flow and Bulk Radial Monodromy}
\label{sec:zero_mode_vs_bulk_monodromy}

The preceding sections place the Virasoro zero-mode structure and the
radial analytic structure of the logarithmic graviton within the same
additive-coordinate geometry. The joint zero-mode evolution depends
nilpotently on $s+\bar s$, while the bulk parametrization identifies
this combination with $2\rho$. We now use this common coordinate
structure to compare the algebraic zero-mode evolution with the
previously established bulk radial monodromy, emphasizing the
different analytic operations associated with the diagonal and
anti-diagonal directions in $(s,\bar s)$ space.

\subsection{Diagonal and anti-diagonal directions}
\label{subsec:jordan_active_inactive}

As shown in Eq.~\eqref{eq:finite_joint_zero_mode_flow}, the nilpotent part of the joint zero-mode evolution depends only on $s+\bar s$. It is therefore useful to introduce the diagonal and anti-diagonal coordinates
\begin{equation}
\sigma:=s+\bar s,
\qquad
\delta:=s-\bar s,
\label{eq:sigma_delta_coordinates}
\end{equation}
so that
\begin{equation}
s=\frac{\sigma+\delta}{2},
\qquad
\bar s=\frac{\sigma-\delta}{2}.
\end{equation}
Using Eq. (\ref{eq:sum_difference_zero_modes}), the exponent of the joint evolution becomes
\begin{equation}
sL_0+\bar s\bar L_0
=
\frac{\sigma}{2}
\left[
(h+\bar h)\mathbf 1+N
\right]
+
\frac{\delta}{2}
(h-\bar h)\mathbf 1.
\label{eq:flow_sigma_delta_decomposition}
\end{equation}
Thus the $\sigma$ direction contains the complete nilpotent
contribution, whereas the $\delta$ direction is purely semisimple.
We shall refer to these, respectively, as the
\emph{diagonal} and \emph{anti-diagonal} directions.

The corresponding directions in the $(s,\bar s)$ coordinate space
are characterized by
\begin{subequations}
\label{eq:active_inactive_summary}
\begin{align}
(\Delta s,\Delta\bar s)
&=(\lambda,\lambda),
&
(\Delta\sigma,\Delta\delta)
&=(2\lambda,0),
\label{eq:active_direction}
\\
(\Delta s,\Delta\bar s)
&=(\lambda,-\lambda),
&
(\Delta\sigma,\Delta\delta)
&=(0,2\lambda).
\label{eq:inactive_direction}
\end{align}
\end{subequations}
The first is the diagonal direction, along which $\sigma$ varies while $\delta$ remains fixed. The second is the anti-diagonal direction, along which $\delta$ varies while $\sigma$ remains fixed.

Under the bulk parametrization established in
Sec.~\ref{sec:additive_coordinates_bulk}, $(\sigma,\delta)=(2 \rho,2i\tau)$.
Thus the diagonal and anti-diagonal directions acquire direct bulk
interpretations as radial variation at fixed $\tau$ and temporal
variation at fixed $\rho$, respectively. In particular, radial variation at fixed $\tau$ follows the diagonal direction and varies the same coordinate $\sigma$ that organizes the radial analytic structure identified in Sec.~\ref{sec:radial_complexification_monodromy}.

The resulting geometry is summarized in
Fig.~\ref{fig:additive_coordinate_geometry}. The diagonal coordinate
$\sigma$ carries the nilpotent zero-mode dependence and, through
$\sigma=2\rho$, also parametrizes the complexified radial direction.
In this description, the finite logarithmic branch points form the
periodic sequence $\sigma_m=(2m+1)\pi i$, while the complementary
coordinate $\delta$, on the physical bulk
section, parametrizes temporal variation at fixed $\rho$. Under the
exponential map $r=e^{\sigma/2}$, the periodic branch sequence
projects to $r=\pm i$, whereas the remaining $r$-plane branch point
$r=0$ corresponds to the asymptotic end
$\operatorname{Re}\sigma\to-\infty$ of the additive-coordinate
covering space.

\begin{figure*}[t]
\centering

\resizebox{\textwidth}{!}{%
\begin{tikzpicture}[
    >=Stealth,
    axis/.style={
        ->,
        black,
        line width=0.75pt
    },
    active/.style={
        ->,
        activeblue,
        line width=1.15pt
    },
    inactive/.style={
        ->,
        inactivered,
        line width=1.15pt
    },
    branchpoint/.style={
        circle,
        fill=black,
        inner sep=2.1pt
    },
    contour/.style={
        contourviolet,
        dashed,
        line width=1.1pt,
        postaction={decorate},
        decoration={
            markings,
            mark=at position 0.18 with {\arrow{>}}
        }
    },
    reference/.style={
        black!55,
        densely dashed,
        line width=0.7pt
    },
    asymptotic/.style={
        ->,
        activeblue,
        line width=1.05pt
    },
    paneltitle/.style={
        font=\small\bfseries,
        align=center
    },
    panelformula/.style={
        font=\small,
        align=center
    },
    annotation/.style={
        font=\scriptsize,
        align=center
    },
    ticklabel/.style={
        font=\scriptsize
    }
]

\begin{scope}[xshift=0cm]

    \draw[black!25]
        (-2.75,-2.75) rectangle (2.75,3.05);

    \node[paneltitle] at (0,2.73)
        {(a) Additive-coordinate directions};

    \node[panelformula] at (1.2,2.00)
        {$(\sigma,\delta)$-plane};

    \draw[axis] (-2.15,0) -- (2.25,0)
        node[right] {$\sigma$};

    \draw[axis] (0,-2.05) -- (0,2.05)
        node[above] {$\delta$};

    \fill (0,0) circle (1.3pt);

    \draw[active] (0,0) -- (1.65,0);
    \draw[active] (0,0) -- (-1.65,0);

    \draw[inactive] (0,0) -- (0,1.55);
    \draw[inactive] (0,0) -- (0,-1.55);

    \node[
        annotation,
        text=activeblue,
        anchor=north west
    ] at (0.45,-0.42)
    {
        Diagonal\\
        $\Delta\sigma\neq0,\quad\Delta\delta=0$\\
        radial variation at fixed $\tau$
    };

    \node[
        annotation,
        text=inactivered,
        anchor=south east
    ] at (-0.35,0.48)
    {
        Anti-diagonal\\
        $\Delta\sigma=0,\quad\Delta\delta\neq0$\\
        temporal variation at fixed $\rho$
    };

\end{scope}

\begin{scope}[xshift=5.75cm]

    \draw[black!25]
        (-2.75,-2.75) rectangle (2.75,3.05);

    \node[paneltitle] at (0,2.73)
        {(b) Branch locus in the $\sigma$-plane};

    \node[panelformula] at (1.2,2.00)
        {$\sigma=s+\bar s=2\rho$};

    \draw[axis] (-2.25,0) -- (2.25,0)
        node[right] {$\operatorname{Re}\sigma$};

    \draw[axis] (0,-2.05) -- (0,2.05)
        node[above] {$\operatorname{Im}\sigma$};

    \fill (0,0) circle (1.2pt);


    \node[branchpoint] at (0,0.70) {};
    \node[ticklabel,anchor=west] at (0.13,0.70)
        {$\pi i$};

    \node[branchpoint] at (0,1.48) {};
    \node[ticklabel,anchor=west] at (0.13,1.48)
        {$3\pi i$};

    \node[branchpoint] at (0,-0.70) {};
    \node[ticklabel,anchor=west] at (0.13,-0.70)
        {$-\pi i$};

    \node[branchpoint] at (0,-1.48) {};
    \node[ticklabel,anchor=west] at (0.13,-1.48)
        {$-3\pi i$};

    \node[annotation] at (0,1.85) {$\vdots$};
    \node[annotation] at (0,-1.86) {$\vdots$};

    \draw[contour]
        (0,0.70) circle[radius=0.46];

    \node[
        annotation,
        text=contourviolet,
        anchor=west
    ] at (0.49,1.05)
        {$\gamma$};


    \draw[asymptotic]
        (-0.15,-0.18) -- (-1.95,-0.18);

    \node[
        annotation,
        text=activeblue,
        anchor=north
    ] at (-1.18,-0.31)
    {
        $\operatorname{Re}\sigma\to-\infty$\\[-1mm]
        $r=e^{\sigma/2}\to0$
    };

    \node[panelformula] at (0,-2.43)
    {
        $\sigma_m=(2m+1)\pi i,\qquad m\in\mathbb Z$
    };

\end{scope}

\begin{scope}[xshift=11.50cm]

    \draw[black!25]
        (-2.75,-2.75) rectangle (2.75,3.05);

    \node[paneltitle] at (0,2.73)
        {(c) Radial geometry in the $r$-plane};

    \node[panelformula] at (1.2,2.00)
        {$r=e^{\sigma/2}=e^\rho$};

    \draw[axis] (-2.25,0) -- (2.25,0)
        node[right] {$\operatorname{Re}r$};

    \draw[axis] (0,-2.05) -- (0,2.05)
        node[above] {$\operatorname{Im}r$};


    \draw[reference]
        (0,0) circle[radius=0.90];

    \node[
        annotation,
        text=black!65,
        anchor=south west
    ] at (0.63,0.60)
        {$|r|=1$};


    \node[branchpoint] at (0,0) {};
    \node[
        ticklabel,
        anchor=north east
    ] at (-0.08,-0.07)
        {$0$};

    \node[branchpoint] at (0,0.90) {};
    \node[
        ticklabel,
        anchor=west
    ] at (0.12,0.90)
        {$+i$};

    \node[branchpoint] at (0,-0.90) {};
    \node[
        ticklabel,
        anchor=west
    ] at (0.12,-0.90)
        {$-i$};


    \draw[contour]
        (0,0) circle[radius=0.47];

    \node[
        annotation,
        text=contourviolet,
        anchor=west
    ] at (0.48,-0.34)
        {$C_{\rm small}$};


    \draw[contour]
        (0,0) circle[radius=1.55];

    \node[
        annotation,
        text=contourviolet,
        anchor=west
    ] at (1.36,1.02)
        {$C_{\infty}$};

    \node[panelformula] at (0,-2.34)
    {
        $\sigma_m\mapsto\pm i,\qquad
        \operatorname{Re}\sigma\to-\infty\mapsto0$
    };

\end{scope}

\end{tikzpicture}%
}

\caption{%
Geometric organization of the complexified additive coordinates.
(a) The diagonal $\sigma$ direction, and the
anti-diagonal $\delta$ direction. Under the bulk
parametrization, these correspond to radial variation at fixed
$\tau$ and temporal variation at fixed $\rho$, respectively.
(b) In the complex $\sigma$-plane, where
$\sigma=s+\bar s=2\rho$, the finite logarithmic branch points form
the periodic sequence
$\sigma_m=(2m+1)\pi i$, with $m\in\mathbb Z$. The asymptotic end
$\operatorname{Re}\sigma\to-\infty$ corresponds under the
exponential map to $r\to0$.
(c) In the complex $r$-plane, the previously established radial
description contains the three finite logarithmic branch points
$r=0,\pm i$. The small contour $C_{\rm small}$ encloses only the
origin, whereas the asymptotic contour $C_\infty$ encloses all three
punctures; the dashed circle indicates $|r|=1$, on which $r=\pm i$
lie. Under $r=e^{\sigma/2}$, the periodic finite branch locus in the
additive coordinate projects to $r=\pm i$, while $r=0$ corresponds
to the asymptotic end $\operatorname{Re}\sigma\to-\infty$.
}
\label{fig:additive_coordinate_geometry}

\end{figure*}


\subsection{Additive-coordinate continuations and radial monodromy}
\label{subsec:opposite_diagonal_continuations}

Having identified the diagonal and anti-diagonal directions,
we now compare two distinct continuations in the additive-coordinate
space. The purpose is to clarify how the bulk radial monodromy is
represented in the same $(s,\bar s)$ geometry as the Virasoro
zero-mode evolution, while distinguishing the analytic continuation
responsible for monodromy from finite zero-mode evolution itself. Consider
first the opposite continuation
\begin{equation}
s\longrightarrow s+2\pi i n,
\qquad
\bar s\longrightarrow\bar s-2\pi i n,
\qquad
n\in\mathbb Z.
\label{eq:opposite_continuation}
\end{equation}
In terms of the diagonal and anti-diagonal coordinates introduced
above, this gives
\begin{equation}
\Delta\sigma=0,
\qquad
\Delta\delta=4\pi i n.
\label{eq:opposite_sigma_delta}
\end{equation}
The continuation therefore lies entirely along the anti-diagonal direction, leaving the nilpotent part of the joint zero-mode evolution unchanged. From Eq.~\eqref{eq:finite_joint_zero_mode_flow},
\begin{equation}
U(s+2\pi i n,\bar s-2\pi i n)
=
e^{-2\pi i n(h-\bar h)}
U(s,\bar s).
\label{eq:opposite_flow_transformation}
\end{equation}
For the critical weights $(h,\bar h)=(2,0)$, the semisimple phase is
unity, and hence
\begin{equation}
U(s+2\pi i n,\bar s-2\pi i n)
=
U(s,\bar s).
\label{eq:critical_opposite_invariance}
\end{equation}
Thus the opposite continuation produces no nontrivial transformation
of the critical logarithmic pair.

Radial analytic continuation has a different structure. At fixed
$\tau$, the bulk parametrization in Eq.~\eqref{eq:bulk_s_sbar}
implies
\begin{equation}
ds=d\bar s=d\rho,
\qquad
d\sigma=2\,d\rho,
\qquad
d\delta=0.
\label{eq:radial_diagonal_continuation}
\end{equation}
It therefore follows the diagonal direction in additive-coordinate space. In contrast to the opposite continuation, $\sigma$ varies along the path and may be analytically continued around the logarithmic branch points identified in Sec.~\ref{subsec:branch_locus_additive}.

The distinction becomes particularly clear for a closed radial
contour $\gamma$. Although its initial and final additive coordinates
coincide,
\begin{equation}
s(1)=s(0),
\qquad
\bar s(1)=\bar s(0),
\label{eq:closed_additive_contour}
\end{equation}
the logarithmic coefficient need not return to its original branch.
Consequently, a nontrivial transformation around such a contour
cannot arise from a net displacement of the additive parameters:
it arises from the multivaluedness of the bulk logarithmic
coefficient along the path of analytic continuation.

As established in
Refs.~\cite{Mvondo-She:2026egr,Mvondo-She:2026eez}, the radial
continuation is determined by the winding data of the logarithmic
branch structure. Denoting the resulting integer monodromy index by
$\nu_\gamma$, the logarithmic coefficient transforms as
\begin{equation}
\Delta_\gamma y
=
-2\pi i\,\nu_\gamma.
\label{eq:previous_bulk_monodromy_y}
\end{equation}
In the Grumiller--Johansson normalization, this induces
\begin{equation}
\psi^L_{\mu\nu}
\longmapsto
\psi^L_{\mu\nu},
\qquad
\psi^{\rm new}_{\mu\nu}
\longmapsto
\psi^{\rm new}_{\mu\nu}
-2\pi i\,\nu_\gamma\psi^L_{\mu\nu}.
\label{eq:previous_bulk_monodromy}
\end{equation}
Equivalently, in terms of the nilpotent operator $N$ introduced in
Sec.~\ref{subsec:joint_zero_modes},
\begin{equation}
M_\gamma
=
\exp(-2\pi i\nu_\gamma N)
=
\mathbf 1-2\pi i\nu_\gamma N.
\label{eq:previous_bulk_monodromy_N}
\end{equation}

The two continuations therefore illustrate different roles of the
same additive-coordinate geometry. Finite zero-mode evolution is
determined by the endpoint in parameter space, whereas bulk radial
monodromy depends on the path of analytic continuation and its
winding around the logarithmic branch structure. The
$(s,\bar s)$ coordinates thus provide a common geometric description
of an endpoint-dependent algebraic evolution and a path-dependent
analytic monodromy.

\subsection{Relation to previous constructions}
\label{subsec:relation_previous_monodromy}

The present construction builds on three related developments in the
logarithmic sector of critical TMG. Bulk radial monodromy of the
logarithmic graviton was established in
Ref.~\cite{Mvondo-She:2026egr} and subsequently extended to the
global descendant sector in Ref.~\cite{Mvondo-She:2026eez}.
Independently, the joint action of the commuting Virasoro zero modes
was formulated in terms of the additive parameters $(s,\bar s)$ in
Ref.~\cite{Mvondo-She:2026vmi}.

The present work connects these developments directly at the level
of the bulk logarithmic solution. The parametrization introduced in
Sec.~\ref{sec:additive_coordinates_bulk} realizes the additive
zero-mode coordinates in terms of the global time and radial
coordinates and expresses the Grumiller--Johansson coefficient
exactly in these variables. In this representation, $s$ governs the
leading complex-linear dependence near the AdS$_3$ boundary, while
$s+\bar s=2\rho$ organizes the radial analytic structure. The finite
periodic branch sequence in the additive coordinate projects to
$r=\pm i$, whereas the remaining $r$-plane branch point $r=0$
corresponds to the asymptotic end
$\operatorname{Re}(s+\bar s)\to-\infty$ of the covering space.
Together, these features provide the additive-coordinate
representation of the singular structure underlying the earlier
radial-monodromy construction.

The significance of the additive-coordinate formulation is therefore
not limited to a rewriting of the logarithmic coefficient. It places
the zero-mode Jordan structure and the bulk radial analytic structure
within the same complex coordinate space and makes their relation
explicit at the level of the logarithmic solution. In particular,
the diagonal coordinate $\sigma=s+\bar s$ simultaneously organizes
the nilpotent zero-mode dependence and the radial analytic geometry
from which the bulk unipotent monodromy arises.

\section{Discussion and Outlook}
\label{sec:discussion}

We have established a direct relation between the additive
coordinates of Virasoro zero-mode evolution and the spacetime
dependence of the logarithmic graviton in critical topologically
massive gravity. With $s=\rho+i\tau$ and
$\bar s=\rho-i\tau$, the Grumiller--Johansson logarithmic
coefficient takes the form
\begin{equation}
y(s,\bar s)
=
-s+\ln2
-\ln\!\left(1+e^{-(s+\bar s)}\right).
\label{eq:discussion_exact_identity}
\end{equation}
Near the AdS$_3$ boundary, $s$ governs the leading complex-linear
dependence of the logarithmic graviton, while the diagonal
combination $s+\bar s=2\rho$ organizes its radial analytic
structure.

Upon complexification of the radial coordinate, the diagonal
combination
\[
\sigma=s+\bar s
\]
contains the periodic sequence of finite logarithmic branch points
$\sigma_m=(2m+1)\pi i$. Under the exponential map
$r=e^{\sigma/2}$, this sequence projects to $r=\pm i$, while the
remaining $r$-plane branch point $r=0$ corresponds to the asymptotic
end $\operatorname{Re}\sigma\to-\infty$ of the additive-coordinate
covering space. The additive formulation therefore gives a geometric
representation of the singular structure underlying the previously
established bulk radial monodromy.

The same diagonal coordinate $\sigma$ controls the nilpotent part of
the joint Virasoro zero-mode evolution. The additive coordinates
therefore provide a common organizing language for the logarithmic
spacetime dependence, the Virasoro Jordan structure, and the radial
analytic geometry associated with bulk monodromy. In this
description, the diagonal and anti-diagonal directions separate the
nilpotent and semisimple components of the zero-mode
parameter space. The two uses of the additive coordinates have
different mathematical origins: finite zero-mode evolution is
determined by additive endpoint data, whereas bulk radial monodromy
is determined by analytic continuation and winding in the
multivalued gravitational solution.

A natural extension is to higher-rank logarithmic sectors and
multicritical three-dimensional massive gravities. It would be
interesting to determine whether analogous additive-coordinate
realizations organize logarithmic and log-square bulk modes, their
branch structures, and the associated higher-rank unipotent
monodromies.

A second direction concerns the global analytic role of the additive
coordinates themselves. In the present rank-two example, the
exponential map distinguishes two features of the radial analytic
structure: the finite periodic branch sequence in the $\sigma$-plane
projects to $r=\pm i$, whereas $r=0$ is represented by an asymptotic
end of the covering space. It would be interesting to determine
whether analogous covering-space descriptions remain useful in
theories with several logarithmic branch loci or more complicated
bulk analytic structures.

\paragraph{Acknowledgements}
The author acknowledges financial support from the Department of Physics at the University of Pretoria.

\clearpage
\bibliographystyle{utphys}
\bibliography{sample}

\end{document}